\documentclass[9pt,twocolumn,twoside]{osajnl}
\usepackage{upgreek}
\usepackage{subfig}
\newsavebox{\measurebox}
\journal{ol} 

\setboolean{shortarticle}{true} 

\title{1.8-THz-wide optical frequency comb emitted from monolithic passively mode-locked semiconductor quantum-well laser}

\author[1,*]{Mu-Chieh Lo}
\author[1]{Robinson Guzm\'an}
\author[1]{Muhsin Ali}
\author[2,3]{Rui Santos}
\author[2,3]{Luc Augustin}
\author[1]{Guillermo Carpintero}

\affil[1]{Universidad Carlos III de Madrid, Av. Universidad 30, 28911 Legan\'es, Spain}
\affil[2]{COBRA Research Institute, Technische Universiteit Eindhoven, 5600 MB Eindhoven, The Netherlands}
\affil[3]{SMART Photonics, Horsten 1, 56l2 AX Eindhoven, The Netherlands}

\affil[*]{Corresponding author: mlo@ing.uc3m.es}

\ociscodes{(140.4050) Mode-locked lasers; (140.5960) Semiconductor lasers; (250.5300) Photonic integrated circuits.}

\begin{abstract}
We report on an optical frequency comb with 14nm ($\sim$1.8 THz) spectral bandwidth at -3 dB level that is generated using a passively mode-locked quantum-well (QW) laser in photonic integrated circuits (PICs) fabricated through an InP generic photonic integration technology platform. This 21.5-GHz colliding-pulse mode-locked laser cavity is defined by on-chip reflectors incorporating intracavity phase modulators followed by an extra-cavity SOA as booster amplifier. A 1.8-THz-wide optical comb spectrum is presented with ultrafast pulse that is 0.35-ps-wide. The radio frequency beat note has a 3-dB linewidth of 450 kHz and 35-dB SNR.
\end{abstract}

\setboolean{displaycopyright}{false}

\begin{document}

\maketitle


Integrated passively mode-locked semiconductor lasers are robust and compact sources for generating coherent optical frequency combs which are of great interests in communications, metrology, spectroscopy and millimeter wave/terahertz (mmW/THz) generation [1]. Unlike other optical frequency comb generation techniques, it does not require any external RF source or optical pump although it can be synchronized to an external RF reference for a lower noise. A passively mode-locked laser is usually a two-section structure, composed of a semiconductor optical amplifier (SOA) as gain section and a saturable absorber (SA) in a Fabry-Pérot cavity [2]. Alternatively, a passively mode-locked laser can be designed in such a way that SAs are placed at certain positions within the resonator to support multiple pulses that meet in the absorbers, which is the colliding-pulse mode locking scheme [3].

\begin{figure}[htbp]
\centering
\includegraphics[width=\linewidth]{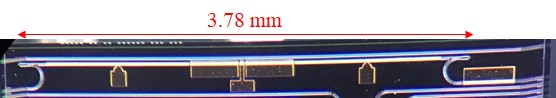}
\includegraphics[width=\linewidth]{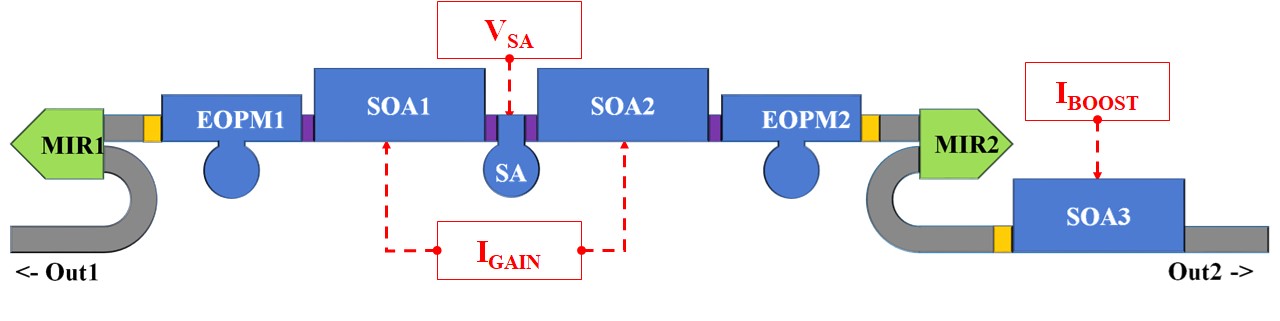}
\caption{(a) Micrograph. (b) Schematic. SA: saturable absorber. SOA: semiconductor optical amplifier. EOPM: electro-optic phase modulator.}
\label{fig01}
\end{figure}

When developed in active-passive integration, passive waveguide components, namely multimode interference couplers/reflectors (MMIs/MIRs), electro-optic phase modulators (EOPMs) are enabled to bring in more functionalities and design freedom [4]. The use of passive waveguides enables SOA and SA lengths to be optimized thus reducing the self-phase modulation in active media [5,6]. Using MMIs and curved waveguides, mode-locked lasers in ring geometry are feasible [7,8]. MIRs allow the on-chip Fabry-Pérot cavity [9], signal acquisition and post-stage processing [10]. In [11-13] it was shown that by introducing a phase shifter-based intracavity gain flattening filter, the optical bandwidth, which is the key characteristic of a comb, can be significantly improved.

Recent studies on the bandwidth of comb generated from a passively mode-locked semiconductor laser have revealed that the performance of quantum-dash (Q-dash) and quantum-dot (QD) active regions is better to that of quantum-well (QW) [14,15]. A QD-based passively mode-locked laser with a 12-nm 3-dB optical bandwidth was presented in [16]. A frequency comb of more than 16 nm was demonstrated from a Q-dash laser in [17]. In [18] a 3-dB optical bandwidth of 11.5 nm was reached with a 20-GHz ring passively mode-locked laser, which was a record value for QW. In [19], the same group proposed another design using the same technology to generate a wide comb with 42-nm 20-dB bandwidth.

In this paper, we present an integrated mode-locked quantum-well laser developed through a generic InP photonic integration platform with a new record bandwidth of 14 nm that is comparably wider than that of a passively mode-locked QD-based laser. The proposed design is a colliding-pulse mode-locked laser in a symmetric arrangement. New features that we have included in this device are symmetrically positioned intracavity EOPMs and an extracavity SOA. The device is demonstrated to generate an extremely broad optical spectrum, that constitutes an optical frequency comb. The phase locking of modes is agreed with the autocorrelation trace, showing ultrafast pulses. The optical comb has a record 3-dB bandwidth of 14 nm (1.8 THz), even wider than that of QD-based passively mode-locked laser.

\begin{figure}[]
  \centering
  \subfloat{\includegraphics[width=0.5\linewidth]{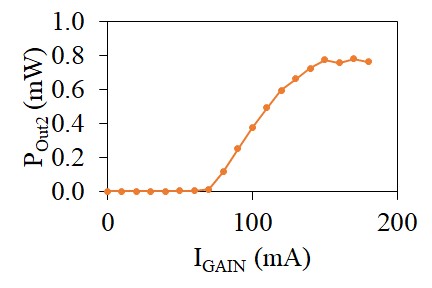}}
  \hfill
  \subfloat{\includegraphics[width=0.5\linewidth]{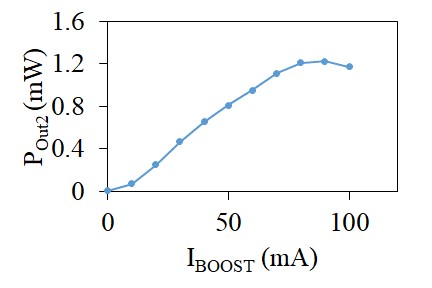}}
\caption{Light-current characteristics (a) P$_{\text{Out2}}$ vs. I$_{\text{GAIN}}$ when V$_{\text{SA}}$ = 0 V and I$_{\text{BOOST}}$ = 50 mA. (b) P$_{\text{Out2}}$ vs. I$_{\text{BOOST}}$ when V$_{\text{SA}}$ = 0 V and I$_{\text{GAIN}}$ = 160 mA.}
\label{fig02}
\end{figure}

Fig. 1(a) depicts the proposed photonic integrated circuit (PIC) microscope photo. Fig. 1(b) presents the schematic. This PIC was developed through a generic approach and fabricated within a multi-project wafer (MPW) run by SMART Photonics [20]. This platform uses ridge waveguides that are 1.5 (deeply etched) and 2.0-$\upmu$m (shallowly etched) wide. The stack has a 500-nm thick InGaAsP Q = 1.25 layer and active waveguides consist of 4 quantum wells. The process design kit (PDK) provides a selection of predefined optoelectronic components as parametric building blocks [21]. The PIC is composed of one SA, two SOAs, two EOPMs, two MIRs and straight/bent passive waveguides in a symmetric geometry with respect to the SA. Without using cleaved facets as reflectors, lithographically defined MIRs assure a reliable and reproducible operation [22], and the total cavity length defined by MIRs is 3.78 mm corresponding to a repetition rate of about 21.5 GHz. This 30-$\upmu$m SA is surround by two 400-$\upmu$m SOAs with two 1200-$\upmu$m-long EOPMs on both sides. Between active components isolations were placed to avoid unwanted current flows. Transitions were inserted between deeply etched and shallowly etched waveguides. The two MIRs have 50\% reflectivity and 50\% transmissivity on two ports. In both MIRs, one port connects to the active components and form a linear resonant cavity. The other port either sends out the pulse train directly to the facet edge through the left 7° angled waveguide (OUT1) or to the other 7° angled waveguide (OUT2) via the 400-$\upmu$m booster amplifier (SOA3). The deeply etched waveguides have a width of 1.5 $\upmu$m and a bending radius of 100 $\upmu$m. Each active component was with a metal pad for electrical contact.

For electrically driving the PIC, the metal pads of PIC were wire-bonded and connected to external DC sources as shown in Fig. 1(b). The PIC was biased with voltage sources (Agilent E3634A Power Supplies) and current sources (Thorlabs PRO8000 LD Controller), denoted as V$_{\text{SA}}$, I$_{\text{GAIN}}$, and I$_{\text{BOOST}}$, respectively. The PIC was mounted on a copper chuck controlled at 16°C with Thorlabs PRO8000 TEC Controller. The optical output signal from OUT2 was coupled to an Oz Optics lensed fiber. The collected optical signal was diagnosed on a Thorlabs power meter (PM), Yokogawa AQ6370B optical spectrum analyzer (OSA), APE PulseCheck autocorrelator (AC) and Anritsu MS2668C electrical spectrum analyzer (ESA). We used an Amonics EDFA (erbium-doped fiber amplifier) to pump the power up to 10 dBm and a polarization controller (PC) prior to the autocorrelator. For the electrical beat tone measurement, a Nortel EDFA with 10 dBm output power and an u2t 40-GHz photodiode (PD) with were employed.

The influence on P$_{\text{Out2}}$ (optical power collected at Out2) of I$_{\text{GAIN}}$ and I$_{\text{BOOST}}$ are shown in Fig. 2(a) and (b), respectively. In Fig. 2(a), as V$_{\text{SA}}$ = 0 V and I$_{\text{BOOST}}$ = 50 mA with I$_{\text{GAIN}}$ varied from 0 to 200 mA, the L-I curve exhibits a typical laser characteristic with threshold current of I$_{\text{GAIN}}$ < 80 mA and the slope efficiency is around 10 mW/A. In Fig. 2(b), as V$_{\text{SA}}$ = 0 V and I$_{\text{GAIN}}$ = 160 mA with I$_{\text{BOOST}}$ varied from 0 to 100 mA, the L-I curve is showing a nearly linear trend with a 15-mW/A slope efficiency and enters saturation region when I$_{\text{BOOST}}$ > 80 mA. The PIC can emit more than 1-mW optical power as the booster amplifier SOA3 is fed with sufficient bias current.

\begin{figure}
\centering
\sbox{\measurebox}{%
  \begin{minipage}[b]{.4\linewidth}
  \subfloat
    
    {\label{fig:figA}\includegraphics[width=\linewidth]{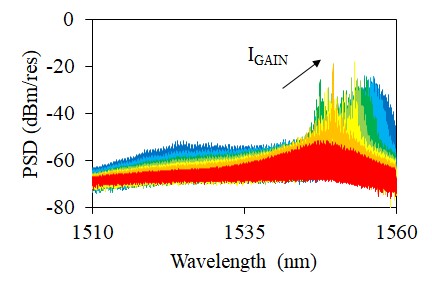}}
    \vfill

\subfloat
  
  {\label{fig:figC}\includegraphics[width=\linewidth]{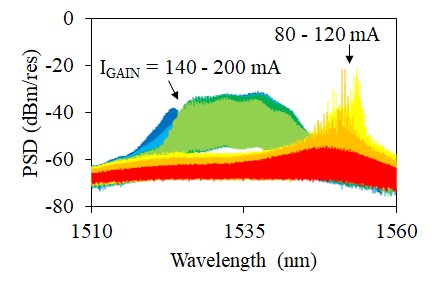}}
  \end{minipage}}
\usebox{\measurebox}\qquad
\begin{minipage}[b][\ht\measurebox][s]{.4\linewidth}
\centering
\subfloat
  
  {\label{fig:figB}\includegraphics[width=\linewidth]{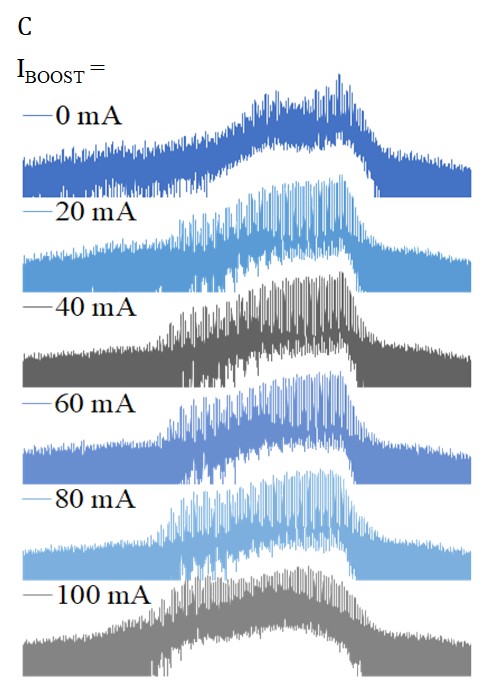}}

\end{minipage}
\caption{(a) Optical spectra at Out2, V$_{\text{SA}}$ = 0 V, I$_{\text{BOOST}}$ = 30 mA, I$_{\text{GAIN}}$ = 80 - 200 mA. (b) Optical spectra at Out2, V$_{\text{SA}}$ = -0.6 V, I$_{\text{BOOST}}$ = 30 mA, I$_{\text{GAIN}}$ = 80 - 200 mA. (c) Optical spectra at Out2, V$_{\text{SA}}$ = -0.6 V, I$_{\text{BOOST}}$ = 0 - 100 mA, I$_{\text{GAIN}}$ = 150 mA.}
\end{figure}

The optical spectra at various bias conditions are plotted in Fig. 3(a) and Fig. 3(b) where the influences of V$_{\text{SA}}$, and I$_{\text{GAIN}}$ are visualized. Under the bias condition of V$_{\text{SA}}$ = 0 V, I$_{\text{BOOST}}$ = 30 mA, and I$_{\text{GAIN}}$ swept from 80 to 200 mA, the optical spectra are shown in Fig. 3(a) while those at V$_{\text{SA}}$ = -0.6 V, I$_{\text{BOOST}}$ = 30 mA, and I$_{\text{GAIN}}$ swept from 80 to 200 mA are shown in Fig. 3(b). The only different parameter between Fig. 3(a) and Fig. 3(b) is V$_{\text{SA}}$. In Fig. 3(a), as I$_{\text{GAIN}}$ increases the lasing wavelength shifts upwards and the profile becomes broader. In Fig. 3(b), similarly the lasing position moves upwards when I$_{\text{GAIN}}$ $\le$ 120 mA. There is a wavelength jump from 1555 nm to 1535 nm, between I$_{\text{GAIN}}$ = 120 mA and I$_{\text{GAIN}}$ = 140 mA. When I$_{\text{GAIN}}$ $\ge$ 140 mA, the wide mode locking spectra are formed about 1535 nm. The optical spectrum is changed not only with V$_{\text{SA}}$ and I$_{\text{GAIN}}$ but also I$_{\text{BOOST}}$. In Fig. 3(c), a series of optical spectra with different I$_{\text{BOOST}}$ are depicted. When V$_{\text{SA}}$ = -0.6 V and I$_{\text{GAIN}}$ = 150 mA, the optical spectrum gradually gets broader and flattened as I$_{\text{BOOST}}$ increases. However, the suppression ratio does not monotonically depend on I$_{\text{BOOST}}$.

\begin{figure}[htbp]
  \centering
  \subfloat{\includegraphics[width=0.5\linewidth]{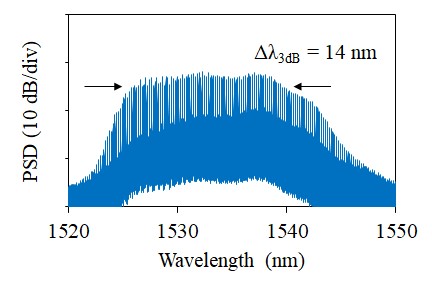}}
  \hfill
  \subfloat{\includegraphics[width=0.5\linewidth]{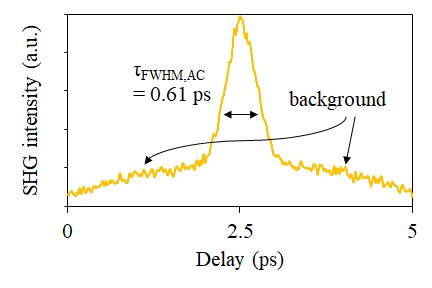}}
\caption{At Out2, VSA = -0.6 V, I$_{\text{GAIN}}$ = 180 mA, I$_{\text{BOOST}}$ = 30 mA. (a) Optical spectrum. Resolution bandwidth: 0.02 nm. 3-dB bandwidth = 14 nm (1.8 THz). Side mode suppression ratio (SMSR) $\sim$ 8 dB. (b) Autocorrelation trace. Full width at half maximum of AC trace $\uptau_{\text{FWHM,AC}}$ = 0.61 ps. Assuming a sech2 pulse shape $\uptau_{\text{FWHM,AC}}$ = 0.35 ps.}
\label{fig04}
\end{figure}

The frequency comb generated from the PIC is demonstrated in Fig. 4(a) with a resolution bandwidth of 0.02 nm. It exhibits an extremely broad spectrum with maximum level at 1537.44 nm, -32.119 dBm/res and the spectral width at 3-dB level is 14 nm. Within the 3-dB bandwidth, it covers 83 comb teeth and the mode spacing is about 0.172 nm corresponding to 21.5 GHz. The side mode suppression ratio (SMSR) is 8 dB which is relatively weak. The wide bandwidth leads to a very short temporal trace, as shown in Fig. 4(b), where a pulse appears at the midpoint of time delay span. The full width at half maximum (FWHM) of AC trace is 0.61 ps. The FWHM of pulse assuming a sech2 shape is 0.35 ps. The time-bandwidth product (TBP) is 1.01, not close enough to the transform limit 0.315, indicating this pulse is chirped and can be further compressed. It should be noted that the pulse is standing on a pedestal. It may be attributed to the complicated intensities with slowly varying envelopes and the non-zero low-frequency amplified component by EDFA or SOA3. Furthermore, the ambiguity in the broad smooth background below the coherence spike probably indicate that the retrieved pulse train is variably spaced. The coherence spike results from the short coherent component of the pulse train, while the background pedestal results from the overall average pulse length [23].

\begin{figure}[bp]
  \centering
  \subfloat{\includegraphics[width=0.5\linewidth]{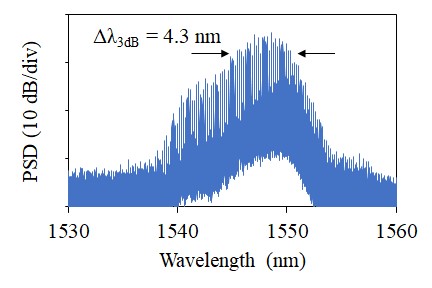}}
  \hfill
  \subfloat{\includegraphics[width=0.5\linewidth]{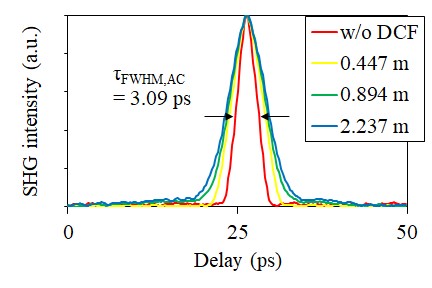}}
\caption{At Out2, VSA = - 1.1 V, I$_{\text{GAIN}}$ = 150 mA, I$_{\text{BOOST}}$ = 50 mA. (a) Optical spectrum. 3-dB bandwidth = 4.3 nm (540 GHz). (b) Autocorrelation trace with different DCF lengths. $\uptau_{\text{FWHM,AC}}$ = 3.09 ps. Assuming a sech2 pulse shape $\uptau_{\text{FWHM,SECH2}}$ = 2.01 ps.}
\label{fig05}
\end{figure}

Another set of AC traces and optical spectrum are shown in Fig.  5(b) and 5(a). The bias condition is V$_{\text{SA}}$ = -1.1 V, I$_{\text{GAIN}}$ = 150 mA, I$_{\text{BOOST}}$ = 50 mA. The 3-dB bandwidth is 4.3 nm (540 GHz), much narrower than that of Fig. 4(a). For evaluating the dispersion of the overall optical path including the PIC under test, lensed fiber and EDFA, three dispersion compensating fibers (DCF) were used, as shown in Fig. 5(b). When no DCF is inserted, the AC FWHM duration is 3.09 ps. Assuming a sech2 shape, the pulse width is 2.01 ps. The duration increases as a longer DCF is applied, confirming the accumulated dispersion is not compensated by the DCFs. Despite this, these AC traces are pedestal-free from which the retrieved pulse train is more stable than that in Fig. 4(b). The TBP is 1.08.

\begin{figure}[t]
  \centering
  \subfloat{\includegraphics[width=0.5\linewidth]{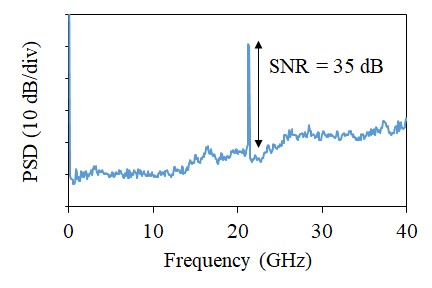}}
  \hfill
  \subfloat{\includegraphics[width=0.5\linewidth]{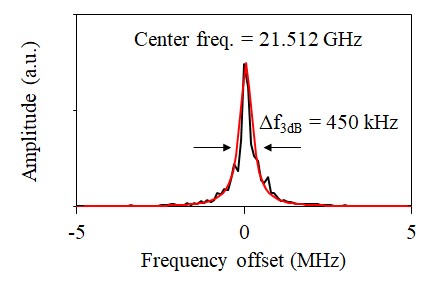}}
\caption{(a) RF spectrum when V$_{\text{SA}}$ = -0.6 V, I$_{\text{GAIN}}$ = 180 mA, I$_{\text{BOOST}}$ = 30 mA. Signal noise ratio (SNR) is 35 dB. (b) RF spectrum with same bias condition. RF Linewidth is 450 kHz.}
\label{fig06}
\end{figure}

The RF spectrum of beat note through the 40 GHz photodiode is demonstrated in Fig. 6. Fig. 6(a) shows the full span up to 40 GHz with 1-MHz resolution bandwidth, while Fig. 6(b) shows the RF tone in detail with 3-kHz resolution bandwidth. In Fig. 6(a), the RF tone signal is at 21.5 GHz and the signal noise ratio (SNR) is about 35 dB. The clean and strong fundamental tone that is 35 dB above any spurious peaks or noise floor indicates that there is little residual amplitude. In Fig. 6(b), the RF tone is observed at 21.512 GHz with a linewidth of 450 kHz.

We have presented a record broad (14 nm $\sim$1.8 THz at 3-dB level) optical frequency comb generated from a passively mode-locked quantum-well (QW) laser. This 21.5-GHz colliding-pulse mode-locked laser with intracavity phase modulators was developed in an InP generic photonic integration technology platform. By inserting a booster amplifier SOA, higher power > 1 mW has been achieved. The autocorrelation trace and RF spectrum have agreed with the mode locking scheme, exhibiting a narrow pulse width of 0.35 ps on a pedestal as well as an RF linewidth of 450 kHz and 35-dB SNR.

\vspace{6pt}
\textbf{Funding.}  This project has received funding from the European Union’s Horizon 2020 research and innovation programme under the Marie Sklodowska-Curie grant agreement No 642355 FiWiN5G, Spanish Ministerio de Economia y Competitividad DiDACTIC project (TEC2013-47753-C3-3-R) and Consejería de Educación, Juventud y Deporte of Comunidad de Madrid DIFRAGEOS project (P2013/ICE-3004). 

\vspace{6pt}
\textbf{Acknowledgment.} We thank Dr. Julien Javaloyes and Prof. Elliott Brown for helpful discussions.

\vspace{6pt}
\textbf{References.}

1.	P. J. Delfyett, S. Gee, M. T. Choi, H. Izadpanah, W. Lee, S. Ozharar, F. Quinlan, and T. Yilmaz, “Optical frequency combs from semiconductor lasers and applications in ultrawideband signal processing and communications,” J. Lightw. Technol. 24(7), 2701-2719 (2006).

2.	J. Javaloyes, and S. Balle, “Mode-Locking in Semiconductor Fabry-Perrot Lasers,” IEEE J. Quantum Electron. 46(7), 1023-1030 (2010).

3.	E. A. Avrutin, J. H. Marsh, E. L. Portnoi, "Monolithic and multi-GigaHertz mode-locked semiconductor lasers: Constructions experiments models and applications," Proc. IEE Optoelectron. 147, 251-278 (2000).

4.	M. Smit, X. Leijtens, E. Bente, J. Van der Tol, H. Ambrosius, D. Robbins, M. Wale, N. Grote, and M. Schell, “Generic foundry model for InP-based photonics,” IET Optoelectron. 5(5), 187-194 (2011).

5.	R. G. M. P. Koumans, R. Van Roijen, "Theory for passive mode-locking in semiconductor laser structures including the effects of self-phase modulation dispersion and pulse collisions," IEEE J. Quantum Electron. 32(3), 478-492 (1996).

6.	K. A. Williams, M. G. Thompson, I. H. White, "Long wavelength monolithic mode-locked diode lasers," New J. Phys. 6, 179 (2004).

7.	M. S. Tahvili, Y. Barbarin, X. J. M. Leijtens, T. de Vries, E. Smalbrugge, J. Bolk, H. P. M. M. Ambrosius, M. K. Smit, and E. A. J. M. Bente, “Directional control of optical power in integrated InP/InGaAsP extended cavity mode-locked ring lasers,” Opt. Lett. 36(13), 2462-2464 (2011).

8.	E. Bente, V. Moskalenko, S. Latkowski, S. Tahvili, L. Augustin and M. Smit, “Monolithically integrated InP-based modelocked ring laser systems,” Proc. SPIE 9134, Semiconductor Lasers and Laser Dynamics VI 91340C (2014).

9.	C. Gordón, R. Guzmán, V. Corral, M.-C. Lo, and G. Carpintero, "On-Chip Multiple Colliding Pulse Mode-Locked Semiconductor Laser," J. Lightw. Technol. 34, 4722-4728 (2016).

10.	M.-C. Lo, R. Guzmán, C. Gordón, and G. Carpintero, "Mode-locked laser with pulse interleavers in a monolithic photonic integrated circuit for millimeter wave and terahertz carrier generation," Opt. Lett. 42, 1532-1535 (2017).

11.	J. S. Parker, A. Bhardwaj, P. R. A. Binetti, Y.-J. Hung, and L. A. Coldren, “Monolithically integrated gain-flattened ring mode-locked laser for comb-line generation,” IEEE Photon. Technol. Lett. 24(2), 131-133 (2012).

12.	V. Moskalenko, J. Javaloyes, S. Balle, M. K. Smit, and E. A. J. M. Bente, “Theoretical Study of Colliding Pulse Passively Mode-Locked Semiconductor Ring Lasers with an Intracavity Mach-Zehnder Modulator,” IEEE J. Quantum Electron. 50(6), 415-422 (2014).

13.	V. Corral, R. Guzmán, C. Gordón, X. J. M. Leijtens, and G. Carpintero, "Optical frequency comb generator based on a monolithically integrated passive mode-locked ring laser with a Mach–Zehnder interferometer," Opt. Lett. 41(9), 1937-1940 (2016).

14.	M. G. Thompson, A. Rae, X. Mo, R. V. Penty, I. H. White, "InGaAs quantum-dot mode-locked laser diodes," IEEE J. Sel. Top. Quantum Electron. 15(3), 661-672 (2009).

15.	G.-H. Duan, A. Shen, A. Akrout, F. V. Dijk, F. Lelarge, F. Pommereau, O. LeGouezigou, J.-G. Provost, H. Gariah, F. Blache, F. Mallecot, K. Merghem, A. Martinez, A. Ramdane, "High performance InP-based quantum dash semiconductor mode-locked lasers for optical communications," Bell Labs Tech. J. 14(3), 63-84 (2009).

16.	R. Rosales, K. Merghem, A. Martinez, A. Akrout, J.-P. Tourrenc, A. Accard, F. Lelarge, and A. Ramdane, “InAs/InP quantum-dot passively mode-locked lasers for 1.55-$\upmu$m applications,” IEEE J. Sel. Top. Quantum Electron. 17(6), 1292 - 1301 (2011).

17.	R. Rosales, S. G. Murdoch, R.T. Watts, K. Merghem, A. Martinez, F. Lelarge, A. Accard, L. P. Barry, and A. Ramdane, "High performance mode locking characteristics of single section quantum dash lasers," Opt. Exp. 20(8), 8649-8657 (2012).

18.	V. Moskalenko, S. Latkowski, S. Tahvili, T. de Vries, M. Smit, and E. Bente, “Record bandwidth and sub-picosecond pulses from a monolithically integrated mode-locked quantum well ring laser,” Opt. Exp. 22(23), 28865-28874 (2014).

19.	V. Moskalenko, J. Koelemeij, K. Williams, and E. Bente, "Study of extra wide coherent optical combs generated by a QW-based integrated passively mode-locked ring laser," Opt. Lett. 42(7), 1428-1431 (2017).

20.	SMART Photonics B. V., http://www.smartphotonics.nl/

21.	Joint European Platform for Photonic Integration of Components and Circuits, http://www.jeppix.eu/

22.	C. Gordón, R. Guzmán, V. Corral, X. Leijtens, and G. Carpintero, “On-chip colliding pulse mode-locked laser diode (OCCP-MLLD) using multimode interference reflectors,” Opt. Exp. 23(11), 14666-14676 (2015).

23.	M. Rhodes, G. Steinmeyer, J. Ratner, and R. Trebino. “Pulse‐shape instabilities and their measurement,” Laser $\&$ Photonics Reviews, 7(4), 557-565 (2013).

\end{document}